\documentclass[10pt,letterpaper]{article}
\usepackage{opex3}
\usepackage{color}
\usepackage[colorlinks=true,urlcolor=blue,linkcolor=black,citecolor=black,breaklinks=true,bookmarks=false]{hyperref}

\newcommand{\unit}[1]{\ensuremath{\;\mathrm{#1}}}
\newcommand{\ket}[1]{\ensuremath{\left| #1 \right\rangle}}

\begin{document}

\title{Effects of photo-neutralization on the emission properties of quantum dots}

\author{Tobias Huber$^{1}$, Ana Predojevi\'{c}$^{1,*}$, Glenn S. Solomon$^{1,2}$, and Gregor Weihs$^{1}$ }
\address{Institut f\"ur Experimentalphysik, Universit\"at Innsbruck, Technikerstrasse 25, 6020 Innsbruck, Austria\\
Joint Quantum Institute, National Institute of Standards and Technology  \& University of Maryland, Gaithersburg, MD 20849, USA
}
\email{ana.predojevic@uibk.ac.at}



\begin{abstract}
In this paper we investigate the coherence properties of a quantum dot under two-photon resonant excitation in combination with an additional photo-neutralization laser. The photo-neutralization increases the efficiency of the excitation process and thus, the brightness of the source, by a factor of approximately 1.5 for biexciton-exciton pairs. This enhancement does not degrade the relevant coherences in the system; neither the single photon coherence time, nor the coherence of the excitation process.
\end{abstract}



\section{Introduction}
Photo-neutralization is the name given to a technique that stabilizes a quantum dot environment using an off-resonant light source~\cite{Nguyen13}. Since it is not possible to grow semiconductors without defects or impurities, a possibility to stabilize the environment is desirable. Photo-neutralization has been shown to be a necessity to achieve resonant excitation~\cite{Metcalfe10, Nguyen12} without electrical bias and, as in our case, it can be used to increase the efficiency of the excitation process. While in previous work continuous wave (CW) resonant excitation of the exciton was studied, we present a study of the effects of photo-neutralization on the pulsed two-photon resonant excitation of the biexciton. This investigation is interesting, because it probes the carrier dynamics in the system at a completely different timescale. Resonant excitation of quantum dots is of great interest \cite{Flagg09, Metcalfe10, Nguyen12, He13, Mueller14}, because of the possibility to coherently control the quantum dots state and to improve the indistinguishability of the emitted photons \cite{Huber15}. Furthermore, resonant pulsed excitation is needed to create single photons with high probability, which are essential for quantum information applications~\cite{Bennett00}. In addition, two-photon resonant excitation is required to generate entangled photon pairs with high probability~\cite{Mueller14}. Therefore, it is important that the photo-neutralization does not degrade the coherence in a pulsed two-photon resonant excitation scheme.
With a simple interference measurement, we demonstrate that photo-neutralization does not negatively influence the coherence of the photon. Nor does it influence the coherence of the excitation process when exciting the system resonantly to the biexciton by exploiting pulsed two-photon absorption. The latter is probed here using time-bin entanglement. Furthermore, time-bin entanglement does not only probe the coherence of the excitation process, but also the posterior free state evolution~\cite{Huber15b}. 

\section{Quantum dots neutralization}
\begin{figure}[ht]
\centering
\includegraphics{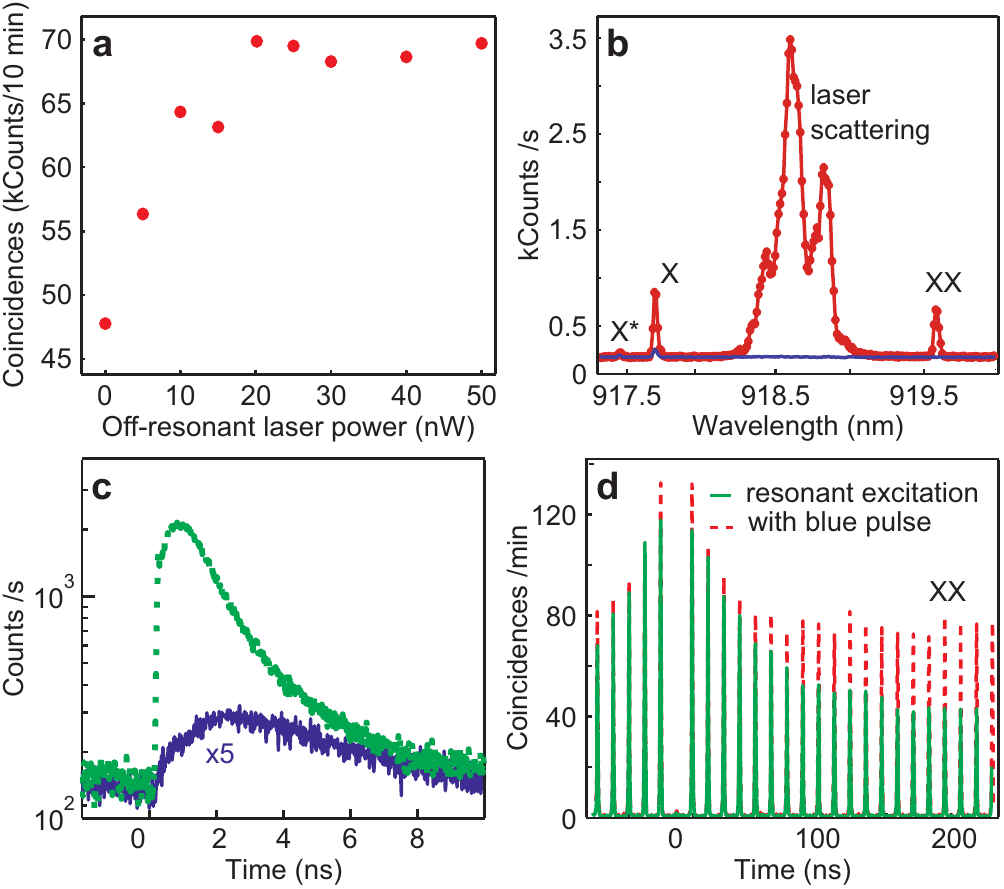}
\caption{(a) Increase of biexciton-exciton coincidence counts obtained from two-photon resonant excitation as a function of off-resonant laser power. (b) In red: emission spectrum under resonant two-photon excitation in the  photo-neutralization regime. In blue: photo-luminescence caused by off-resonant laser alone. (c) Emission probability of the exciton photon vs. time. In green dotted: only resonant two-photon excitation, in blue: excitation caused by the off-resonant laser alone. (d) Autocorrelation measurement performed on biexciton photons. The data plotted in red dashed was obtained in photo-neutralization regime.}
\label{fig:blinky}
\end{figure}

Recently, Nguyen et al. performed studies \cite{Nguyen12, Nguyen13} of emission quenching in quantum dots under cw resonant excitation. They showed that the origin of this phenomenon lies in the Coulomb blockade caused by a residual hole. Additionally, they demonstrated that by applying a weak off-resonant laser one can photo-neutralize the quantum dot device and enable resonant excitation. 
\begin{figure}[ht]
\centering
\includegraphics{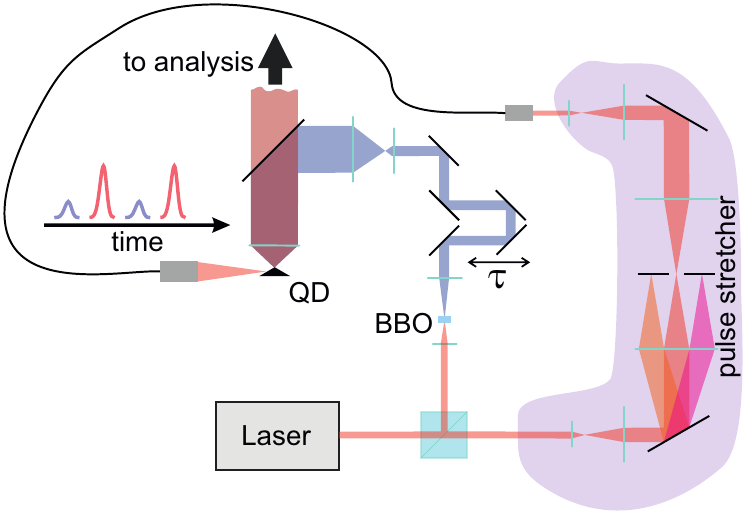}
\caption{Schematics of the excitation part of the setup. A pulse stretcher is used to modify the pulse length of the two-photon excitation laser to 4\unit{ps}. A part of the beam is sampled and frequency doubled with a BBO crystal. This blue light is used as photo-neutralization laser. A delay stage introduces an additional delay $\tau$ to control the relative timing between the resonant laser and the photo-neutralization laser.}
\label{fig:setup}
\end{figure}
While in our samples we do not observe such a high level of emission quenching as shown in Refs.~\cite{Nguyen12, Nguyen13}, we do observe blinking. This blinking manifests itself in an exponential decay in the autocorrelation measurements (in red in Fig.~\ref{fig:blinky}(d)). Additionally, we observe some trion emission (Fig.~\ref{fig:blinky}(b)) irrespective of the method of excitation, i.e. above-band excitation and two-photon resonant excitation. Trion emission indicates the presence of a residual charge in the quantum dot. A defect in the vicinity of the quantum dot acts as a hole or electron reservoir \cite{Nguyen12, Nguyen13}, from where it can be captured by the quantum dot.
Another possible explanation for the blinking effect could be an energy shift of the biexciton state caused by the electric field of the trapped charges in the vicinity of the quantum dot, through the quantum confined Stark effect \cite{Abbarchi08}. Since the shift of the level is only a few GHz and the laser linewidth is 160\unit{GHz}, we do not expect to see such a strong blinking behaviour~\cite{Jayakumar13}. Also, the existence of the trion under two-photon resonant excitation cannot be explained with this effect. Therefore, we think that a defect close to the quantum dot has to be the origin of the observed blinking.

\section{Methods}

The sample was held at a temperature of 4.8\unit{K}. It contains low-density, self-assembled InAs quantum dots embedded in a planar micro-cavity. The unintentional background doping is likely in the $10^{14}\unit{cm^{-3}}$ range. The excitation light was derived from a pulsed, tunable 82\unit{MHz} Ti:sapphire laser with a pulse length of 2\unit{ps}. Using a pulse-stretcher, the length of the pulses was adjusted to 4\unit{ps}. The laser wavelength was 918.7\unit{nm}, which is half the energy of biexciton plus exciton emission energy (see Fig.~\ref{fig:scheme}(b)).  For photo-neutralization we used pulsed blue light obtained from  second harmonic generation of the two-photon resonant laser in a 3\unit{mm} long BBO crystal. Using a delay stage, we could time the arrival of the photo-neutralization pulse to be in between two subsequent excitation pulses. The setup is schematically shown in Fig.~\ref{fig:setup}.

\section{Results}
To fully describe the influence of the off-resonant laser on the system, we performed various experiments. These include, measurements of the photon generation efficiency, influence on  the spectrum, influence on the lifetime, influence on the multiphoton contribution, influence on the coherence of the emitted photons and the influence on the coherence of the excitation process. Fig.~\ref{fig:blinky}(a) shows the effect of the photo-neutralization laser on the photon generation efficiency. Here, we increased the power of off-resonant laser while keeping the two-photon resonant laser at a fixed power level. The photon generation efficiency increases dramatically up to about 20\unit{nW} of added photo-neutralization laser, where the photon generation efficiency reaches a plateau, see Fig.~\ref{fig:blinky}(a).  Adding the photo-neutralization laser to the two-photon resonant laser increases the photon generation efficiency of the source by a factor of 1.49(1), where we defined the photon generation efficiency as the number of detected biexciton-exciton pairs.
\begin{figure}[ht]
\centering
\includegraphics{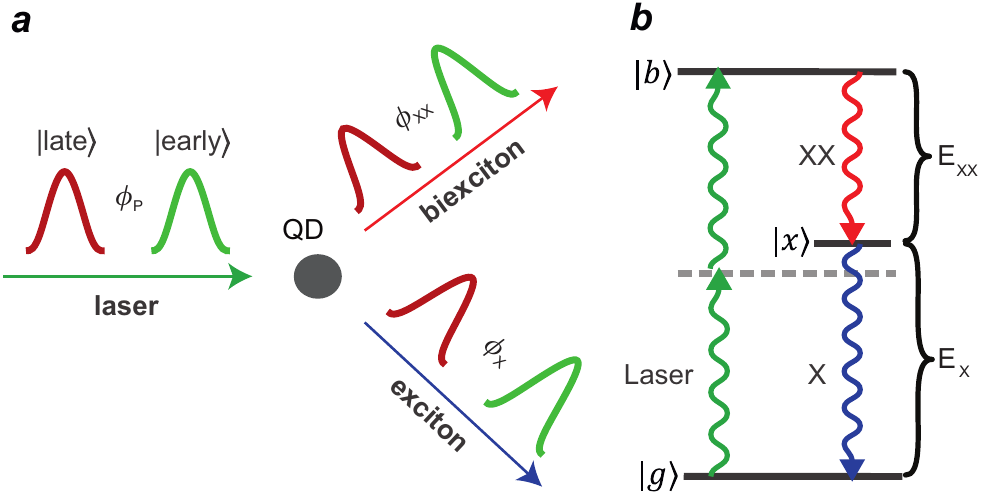}
\caption{(a) Time-bin excitation scheme. The quantum dot is excited with two pulses \ket{early} and \ket{late} with a defined phase $\phi_P$ between them. 
The resulting state is given by Eq. (\ref{eq:state}) (b) Energy scheme. A two-photon process drives the system coherently from the ground state \ket{g} to the biexciton state \ket{b}. The decay proceeds via the intermediate exciton level \ket{x}, resulting in a photon cascade. }
\label{fig:scheme}
\end{figure}
A photoluminescence spectrum when exciting only with the off-resonant blue laser compared to the emission spectrum under two-photon resonant excitation with the off-resonant laser added can be seen in Fig.~\ref{fig:blinky}(b).  The photoluminescence caused by the off-resonant blue laser alone is small and the biexciton-exciton pair rate created by the photo-neutralization laser alone is zero (not shown).
Further, a time resolved measurement of the exciton emission shows that the exciton decay, when excited with the photo-neutralization laser, is much slower (longer rise time) than the normal exciton decay (see Fig.~\ref{fig:blinky}(c)). The decay in resonant two-photon excitation is depicted as green dots. The decay obtained from the off-resonant blue pulse, used for the photo-neutralization, without the resonant excitation is depicted as blue solid line. For the photo-neutralization laser only, the fitted function $A\cdot (1-e^{-x/t_3})\cdot e^{-x/t_1}+c$ resulted in a level filling time $t_3=4.8(2)\unit{ns}$ and a decay time $t_1=2.3(1))\unit{ns}$, compared to $t_1=1.0(2)\unit{ns}$ in two-photon resonant excitation. 
The slow filling time is caused by the large energy difference between the blue laser and the exciton resonance so that carriers must radiatively decay through a series of states before occupying the ground state. The origin of the significant change of the lifetime ($t_1$) of the exciton is unclear.\\
To investigate the influence of photo-neutralization to the emitted photons, we measured the auto-correlation of the biexciton photon with and without the photo-neutralization laser.  For this measurement, the emission of the biexciton photon is sent onto a 50:50 beamsplitter, whose outputs are detected by an avalanche photo diode each. The detection of a photon at one of the detectors defines the zero point. The time delay to a photon detected at the other detector is plotted as a histogram.
The auto-correlation is shown in Fig.~\ref{fig:blinky}(d). The amount of signal at time zero compared to the signal at other times gives a measurement for the multi-photon contribution of the source. It can be seen that there is no measurable increase in the multi-photon contribution caused by the photo-neutralization laser.  Further, in the same figure, it can be seen that the photo-neutralization laser is reducing the blinking. 
The detection of the second photon in the auto-correlation measurement in Fig.~\ref{fig:blinky}(d) is more likely to occur within a short timescale after detection of a (start) photon, which  manifests in an exponential decay of the envelope of the autocorrelation function. This can be caused by a defect in the vicinity of the quantum dot which is charging it. The additional charge prohibits the excitation of the biexciton state. This is the blinking effect.
\\
 \\
Most importantly, we wanted to check the influence of the photo-neutralization laser on the coherences of the system.  We first investigate the influence on the free evolution of the state, and second investigate  the influence on the coherence of the excitation process. The latter can be probed utilizing a time-bin entanglement scheme. The observed amount of time-bin entanglement is an indirect measurement of the coherence between the ground state (the vacuum state) and the biexciton state as we will describe in detail later.

\begin{figure}[ht]
\centering
\includegraphics{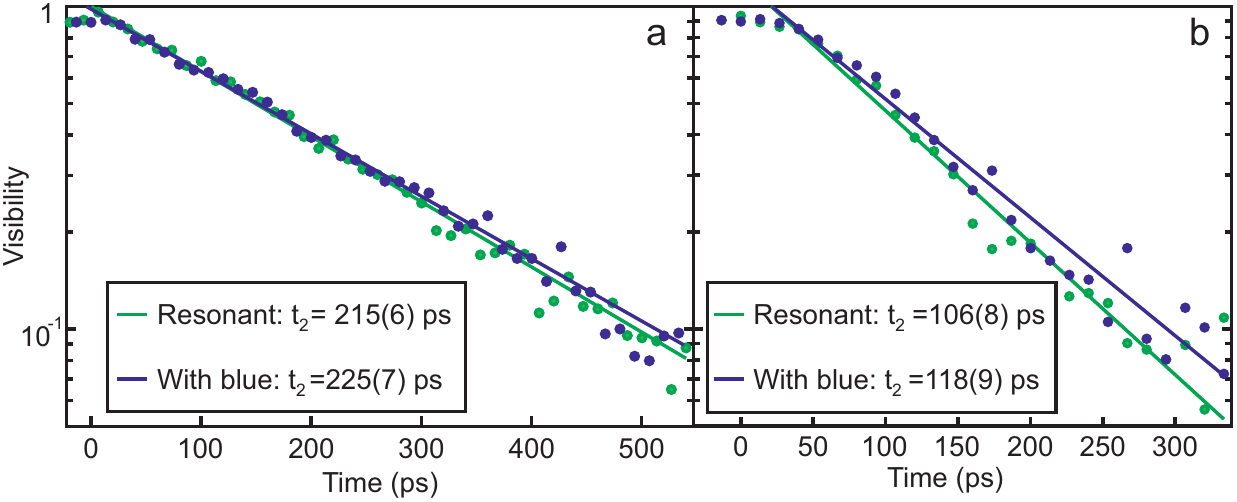}
\caption{(a) Coherence of the XX photon excited with two-photon resonant excitation with and without the off-resonant blue photo-neutralization laser. (b) Same as (a) for the  X photon.}
\label{fig:coherence}
\end{figure}

To measure the coherence time of the photons ($g^{(1)}$), we utilized an unbalanced Michelson interferometer with one variable arm to introduce a time delay, see Fig.~\ref{fig:setup}. The fixed arm is mounted on a piezo translation stage to vary the phase of the interferometer. The visibility of the fringe contrast is plotted as a function of time delay in Fig.~\ref{fig:coherence}. The data shows a slight improvement of the coherences using photo-neutralization (with 20\unit{nW} blue laser), both for biexciton and exciton photon. This could come from a stabilization of the environment due to the additionally created charge carriers.

To explain how time-bin entanglement was utilized to measure the coherence of the excitation process, we introduce the scheme of time-bin entanglement first.
Time-bin entanglement is entanglement of photons in two temporal modes (time-bins): early and late. 
Such a scheme is depicted in Fig.~\ref{fig:scheme}(a). A laser pulse is split into two pulses early and late in an interferometer. These two pump pulses are used to create a pair of photons either in the early or in the late pulse.  
The early and late pulse need to have a fixed phase relation, because this phase is transfered directly onto the entangled state (see Eq.~(\ref{eq:state})).
In quantum dots resonant excitation is used to coherently drive the system from ground to excited state (see Fig.~\ref{fig:scheme}(b)). The coherent excitation transfers the phase of the laser to the quantum dot, which is encoding the pump phase onto the entangled state. The generated entangled state is of the form

\begin{equation}
|\Phi\rangle=\frac{1}{\sqrt2}(|early\rangle_{1}|early\rangle_{2}+e^{i\phi_{P}}|late\rangle_{1}|late\rangle_{2}),
\label{eq:state}
\end{equation}

where $\phi_{P}$ is the phase of the pump interferometer and $|early\rangle$ ($|late\rangle$) denote photons generated in an early (late) time-bin. 1 (2) stands for the exciton (biexciton) recombination photon in a quantum dot.
\\

\begin{figure}[ht]
\includegraphics{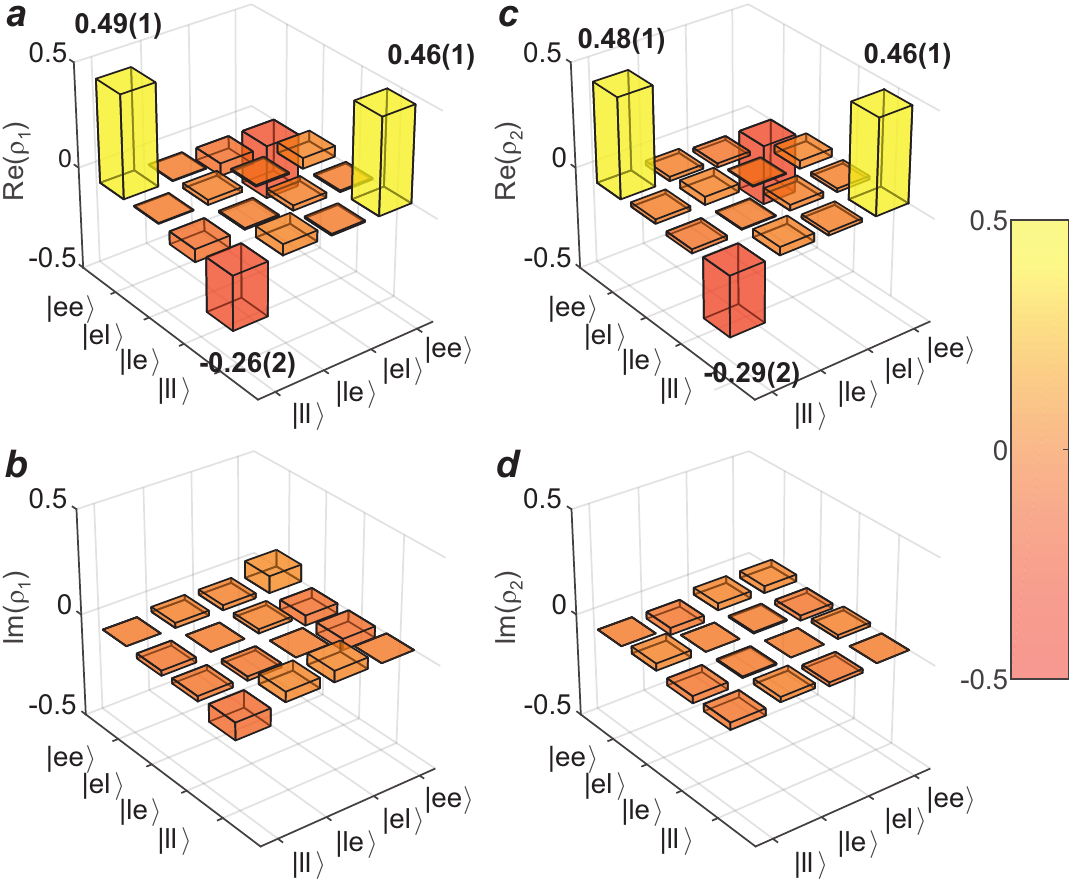}
\centering
\caption{Reconstructed density matrices for the time-bin entangled photon pair. \ket{e} is short for \ket{early} and \ket{l} is short for \ket{late}. Indices are not shown for brevity. (a) and (b) Density matrix for two-photon resonant excitation, (c) and (d) with added photo-neutralization laser.}
\label{fig:matrix}
\end{figure}

To characterize the time-bin entanglement we performed state tomography \cite{James01, Takesue09}. The fidelity of the reconstructed two-photon density matrix with the maximally entangled \ket{\Phi^-} Bell state was found to be F=0.74(3) while the concurrence was C=0.51(7). The same measurement made with photo-neutralization resulted in a fidelity of F=0.76(3) while the concurrence was measured to be C=0.54(6). For both measurements the excitation probability was $6\%$. This resulted in 15.9 detected pairs/s with photo-neutralization instead of 10.7 pairs/s without it. The reconstructed density matrices for the two measurements are shown in Fig.~\ref{fig:matrix}. Each measurement point was averaged over 10\unit{min}. The density matrix was reconstructed using the maximum-likelihood estimation method. In order to obtain the measurement errors we performed a 100 run Monte Carlo simulation of the data with a Poissonian noise model applied to the measured values.

\section{Conclusion}
We investigated the method of photo-neutralization in the regime of pulsed two-photon resonant excitation of the biexciton. Thereby, we showed that optical neutralization of the quantum dot neither degrade the coherence of the photons, nor the amount of time-bin entanglement, which can be used as a measure for the coherence of the excitation process. Instead, photo-neutralization can significantly increase the efficiency of the source.

\section*{Acknowledgments}
This work was funded by the European Research Council (project EnSeNa, no. 257531). T. H. thanks the Austrian Academy of Sciences (\"OAW) for receiving a DOC Fellowship. A. P. would like to thank Austrian Science Fund (FWF) for support provided through Elise Richter Fellowship V-375. G.S.S. acknowledges partial support from the NSF PFC@JQI, and from Fulbright Austria - Austrian American Educational Commission through the Fulbright-University of Innsbruck Visiting Scholar program.

\end{document}